\documentclass[letterpaper,11pt]{article}

\usepackage{jheppub}
\usepackage{hyperref}
\usepackage{epsfig,amssymb,amsmath,psfrag,epstopdf,color,float}
\usepackage{graphicx}
\allowdisplaybreaks

\def\beq{\begin{equation}}
\def\eeq{\end{equation}}
\def\bsp#1\esp{\begin{split}#1\end{split}}
\newcommand{\be}{\begin{equation}}
\newcommand{\ee}{\end{equation}}
\newcommand{\bea}{\begin{eqnarray}}
\newcommand{\eea}{\end{eqnarray}}

\preprint{}

\title{Probing light-quark Yukawa couplings via hadronic event shapes at lepton colliders}

\author{Jun Gao}

\affiliation{INPAC, Shanghai Key Laboratory for Particle Physics and Cosmology, School of
Physics and Astronomy, Shanghai Jiao-Tong University, Shanghai 200240, China}

\emailAdd{jung49@sjtu.edu.cn}

\abstract{We propose a novel idea for probing the Higgs boson couplings through
the measurement of hadronic event shape distributions in the decay of the Higgs boson at lepton
colliders. The method provides a unique test of the Higgs boson couplings and of
QCD effects in the decay of the Higgs boson. It can be used to
probe the Yukawa couplings of the light quarks and to further test the
mechanism of electroweak symmetry breaking.
From a case study for the proposed Circular Electron-Positron
Collider, assuming a hypothesis of SM-like theory, light-quark
couplings with a strength greater than 9\% of the bottom-quark Yukawa
coupling in the standard model can be excluded.}     

\keywords{Higgs boson, lepton collider, QCD}

\begin{document} 

\maketitle

\section{Introduction}

The successful operation of the CERN Large Hadron Collider (LHC) and
the ATLAS and CMS experiments have
led to the discovery of the Higgs boson, the final piece of the
standard model (SM)~\cite{Chatrchyan:2012ufa,Aad:2012tfa} of particle physics.
Future high precision experimental investigations on the couplings of the Higgs boson
are required for a refined understanding of the nature of electroweak symmetry
breaking and for searches for possible new physics beyond the SM.
Higgs boson couplings can be measured to percent level precision
at future lepton colliders, e.g., the International Linear Collider~\cite{1310.8361}
and the Circular Electron-Positron Collider (CEPC)~\cite{CEPC-SPPCStudyGroup:2015csa}, or
with less precision at the high luminosity run of the LHC (HL-LHC)~\cite{1310.8361}.
In addition to high precision, $e^+e^-$ colliders provide direct access to
all possible decay channels of the Higgs boson, including invisible decays,
in a clean environment. They can also measure the total width of the Higgs
boson in a model-independent way.  

An important prediction for the SM Higgs boson is that the couplings to other
SM particles are proportional to their mass. It will be essential to test
this relation experimentally. In the SM the Yukawa couplings of the
Higgs boson to light quarks $q$ ($u$, $d$, or $s$) are negligibly small due smallness
of their mass. There have been, however, theoretical models that have predicted enhanced
light-quark Yukawa couplings~\cite{0804.1753,1504.04022}. Experimentally,
if such an enhanced-coupling scenario is observed, it will must
indicate the presence of new physics; the quarks also receive masses from
sources other than the Higgs boson in order to maintain a relatively small mass.
However, a direct measurement of light-quark Yukawa couplings is impossible
at hadron colliders due to the huge QCD backgrounds for hadronic decays of
the Higgs boson. Indirect constraints can be obtained based on different
kinematic distributions induced by gluon and quark production
mechanisms~\cite{1308.5453,1606.09253,1606.09621} or
through rare decays of the Higgs
boson~\cite{Bodwin:2013gca,Kagan:2014ila,Zhou:2015wra,Koenig:2015pha,Perez:2015lra,Chisholm:2016fzg}.

At lepton colliders, the main measurement difficulty is separation of the
$q\bar q$ decay channel from the loop-induced gluon channel, both of which generate similar
final states of two untagged jets ($jj$).  
In this work, we propose a novel idea of using hadronic event shape observables from the Higgs
boson decays to separate $q\bar q$ from $gg$ channels and to measure the light-quark
Yukawa couplings at lepton colliders.
Another possibility for lepton colliders involves utilizing discrimination of quark
jets and gluon jets~\cite{1605.04692}. We leave this for future investigations. The
idea is motivated by the measurement of the QCD coupling constant at LEP from
hadronic event shape distributions. 
\footnote{Event shapes have been employed to study the spin and $CP$ property of the Higgs
boson at the LHC~\cite{Englert:2012ct,Englert:2013opa}.}
Intuitively, in that case the next-to-leading order
QCD corrections, $\sim {\mathcal O}(\alpha_s)$, generate the distribution in
three-jet region. A change of $\alpha_s$ can induce changes of the event shape
distributions, e.g., the position and height of the peak. Similarly, in the case of
the Higgs boson decay, the real radiation is of ${\mathcal O}(C_X\alpha_s)$,
where $C_X$ is the QCD color factor, i.e. $C_A=3$ for decay to gluons and $C_F=4/3$ for
decay to quarks. Thus, a measurement of event shape distributions can reveal
the average color factor and the ratio of decay branching ratios (BR) of
the gluon and the quark channel.    
         
In the remaining paragraphs we demonstrate theoretically how the distributions
differ for quark and gluon channels, and we consider a scenario of the CEPC
and demonstrate a precision of $<1\%$ can be achieved on the
measurement of the decay BR to light quarks. \\  

\section{Event shapes}

There have been 6 major observables of hadronic event shapes measured at LEP and
used for the extraction of $\alpha_s(M_Z)$, including thrust $T$ (or $\tau=1-T$),
heavy hemisphere mass $M_H$, $C$ parameter, total hemisphere
broadening $B_T$, wide hemisphere broadening $B_W$, and the
Durham 2 to 3-jet transition parameter $y^D_{23}$~\cite{Abbiendi:2004qz,Heister:2003aj}.
For example, the thrust is defined as
\begin{equation}
T= \max_{\vec n}\left(\frac{\sum_i|p_i\cdot \vec{n}|}{\sum_i|p_i|}\right),
\end{equation}
where $p_i$ is the three-momentum of particle $i$ and the summation runs over all
measured particles. One advantage of the global event-shape observables is that
their distributions can be calculated systematically in perturbative
QCD\cite{Banfi:2014sua,Banfi:2016zlc}.
In case of two-body hadronic decay, at the leading order (LO),
the thrust distribution is a $\delta$ function at $\tau=0$. Finite thrust values
are generated through high-order QCD radiations. Soft and collinear emissions
introduce large logarithmic contributions $\sim \alpha_s^n \ln \tau^{2n-1}/\tau$
at small-$\tau$, the deep two-jet region. They must be resummed to all orders
in QCD to make reliable predictions, e.g., the state of art Next-to-Next-to-Next-to-leading
logarithmic ($\rm N^3LL$) resummation~\cite{0803.0342,Abbate:2010xh,1411.6633} for
$Z/\gamma^*\rightarrow q\bar q$ in the extraction of $\alpha_s(M_Z)$.
Meanwhile, in the three-jet region the resummed results can be further
matched with the fixed-order results, e.g., the Next-to-Next-to-leading
order (NNLO) calculation for $Z/\gamma^*\rightarrow 3\ jets$ production~\cite{0707.1285,1606.03453}.
Usually, for calculations done at parton level, a correction factor due to
hadronization effects needs to be applied when comparing to experimental data,
which can be estimated through various event generators~\cite{Lonnblad:1992tz,0710.3820,0803.0883,0811.4622}.

To our best knowledge, no predictions at comparable precision exist
for hadronic decays of the Higgs boson, although most of the ingredients
are already available. Predictions at $\rm N^3LL+$NNLO level for the Higgs boson
are expected in near future. In this study, we calculate the
event shape distributions using the MC event generator Sherpa 2.2~\cite{0811.4622}
with the effective coupling approach of the Higgs boson. We
use the CKKW scheme~\cite{hep-ph/0109231}, matching parton showers with tree-level
matrix elements with up to three jets, which is effectively partial next-to-leading-logarithmic
and leading-order accuracy. The hadronization corrections are included
automatically in Sherpa simulation through hadronization models and decays of hadrons.

\begin{figure}[!h]
  \begin{center}
  \includegraphics[width=0.45\textwidth]{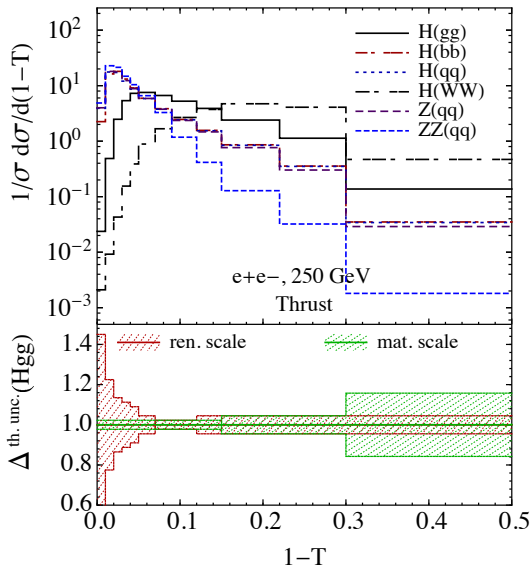}
  \end{center}
  \vspace{-2ex}
  \caption{\label{fig:thrust}
   Normalized distributions of thrust in hadronic decays of the Higgs boson,
   in $e^+e^-\rightarrow q\bar q$ with a CMS
   energy of 125 GeV and in $e^+e^-\rightarrow Z q\bar q$ with a CMS energy of 250 GeV.
   In the later two cases the thrust is calculated in the hadronic CMS frame. 
   The lower panel shows the relative
   theoretical uncertainties of the normalized distribution for $H\rightarrow gg$,
   due to the variations of renormalization and matching scales.}
\end{figure}

Fig.~\ref{fig:thrust} shows the normalized distribution of the
variable thrust for several different hadronic decay channels of the Higgs boson, including
$gg$, $q\bar q$, $b\bar b$, and $W(q\bar q)W^*(q\bar q)$. We also plot the
distribution in the hadronic center-of-mass (CMS) frame
for $e^+e^-\rightarrow Z^*/\gamma^* \rightarrow q\bar q$ with a CMS energy of
125 GeV and $e^+e^-\rightarrow Zq\bar q$ with a CMS energy of 250 GeV
and requiring a recoil mass of 125 GeV of the $Z$ boson as comparisons. The
distribution peaks at $\tau\sim 0.02$ for light-quark decay channel. The
peak shifts to $\tau\sim 0.05$ for the gluon channel, corresponding to
a scaling of roughly $C_A/C_F$. The distribution is much broader for the
gluon case due to
the stronger QCD radiation. The distribution for the $b\bar b$ channel is very close
to the $q\bar q$ case, except at very small $\tau$, where the mass and hadronization
effects become important. For the $WW^*$ channel there exist already four
quarks at LO and the distribution is concentrated in the large-$\tau$ region.
The distribution for $q\bar q$ from $Z^*/\gamma^*$ differs from
that for the Higgs boson in the
three-jet region because of the different spin. The distribution for $q\bar q$
in $Zq\bar q$ production has a slightly higher peak than the case of $Z^*/\gamma^*$
mostly due to the different hard radiation patterns and different
tunings of parton showering. 

In the lower panel of Fig.~\ref{fig:thrust}, we plot the estimated theoretical
uncertainties of the normalized thrust distribution for the decay to
gluons. 
Theoretical uncertainties due to the truncation
of the perturbation series are conventionally estimated through QCD scale
variations. 
These include variations due to the change of the renormalization
scale and the matching scale~\cite{Jones:2003yv}. 
The latter variation mostly affects the distribution in
the large-$\tau$ region.
As one includes higher-order resummation and
fixed-order matching contributions, the scale variations will decrease.
We assume a $\rm N^3LL+$NNLO calculation for the Higgs boson decay to gluons will
be available and estimate the scale variations based on the
calculation for $Z/\gamma^*$~\cite{0906.3436,0803.0342} using a scaling factor
of $C_A/C_F$. 
Since the distribution is normalized, the uncertainties are
small in the peak region.
The uncertainty due to the $\alpha_s(M_Z)$ input is negligible if the
world average~\cite{Beringer:1900zz} is used. 

\begin{figure}[!h]
  \begin{center}
  \includegraphics[width=0.4\textwidth]{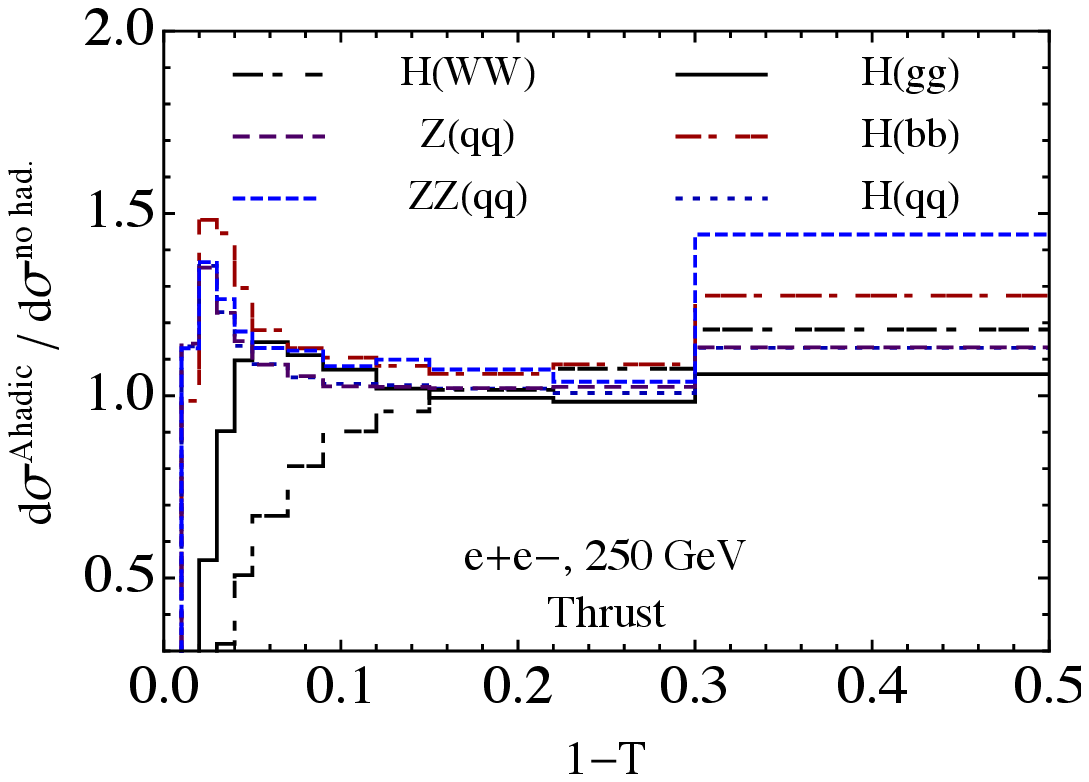}
  \hspace{0.4in}
  \includegraphics[width=0.4\textwidth]{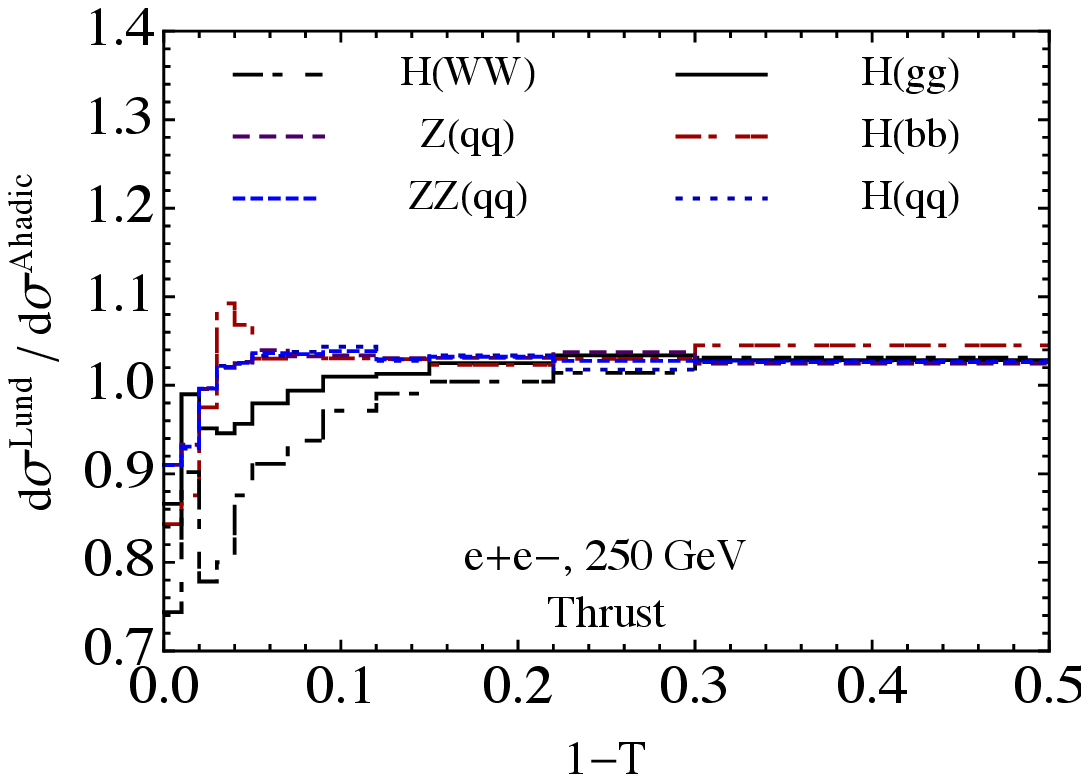}
  \end{center}
  \vspace{-2ex}
  \caption{\label{fig:hadro}
   Ratios of normalized distributions of thrust,
   left: predictions with AHADIC++ hadronization model to without hadronization corrections;
   right: predictions with Lund hadronization model to with AHADIC++ hadronization model.}
\end{figure}

There are also uncertainties due to the hadronization model used.
Sherpa uses a cluster fragmentation model implemented in AHADIC++~\cite{Winter:2003tt}
by default with which the results in Fig.~\ref{fig:thrust} are simulated.
In Fig.~\ref{fig:hadro} the left plot shows the size of hadronization corrections
by taking ratio of the normalized distributions with and without turning on
the hadronization module in Sherpa.
We can see roughly three patterns of the hadronization corrections in Fig.~\ref{fig:hadro}.
All distributions initiated from $q\bar q$ and $b\bar b$ final states receive
similar corrections.
The distributions are enhanced by more than 30\% around the peak region and
are greatly reduced when thrust goes to one as a balance.
That can be understood since the hadronization effects will distribute energies
away form the jet axis.
Shape of hadronization corrections for distribution of $H(gg)$ is much broader and shifted
to the right side as comparing to $q\bar q$ cases.
Lastly the distribution of $H(WW)$ is further suppressed at small $\tau$ region
by hadronization corrections.
To estimate uncertainties due to the hadronization corrections we recalculate
all the distributions with the alternative hadronization model in Sherpa by
linking to the Lund string fragmentation in PYTHIA 6.4~\cite{Sjostrand:2006za}.
We plot ratios of predictions from the two different hadronization models in the right
plot in Fig.~\ref{fig:hadro}.       
The differences can be large for thrust greater than 0.9, about +10(-5)\% for
$H(b\bar b)$($H(gg)$) at thrust $\sim 0.95$, and become even larger when entering
fully non-perturbative dominant region.
We can take above differences as the size of hadronization
uncertainties which are summarized in Table~\ref{tab:hadro} for two representative
bins of $\tau$.
All the $q\bar q$ cases have small uncertainties in the peak region.
Relative signs in Table~\ref{tab:hadro} indicate the uncertainties in
different bins are either fully correlated or anti-correlated.
Though hadronization uncertainties of all channels discussed are derived
from the same models, we decorrelate the uncertainties of different channels
to be conservative, which are described by individual nuisance parameters.  
Below, we will discuss the possibility of measuring the distributions discussed above
at a lepton collider and the sensitivity of these measurements to
the light-quark Yukawa couplings. \\

\begin{table}[h!]
\centering
\begin{tabular}{c|cccccc}
\hline
had. unc. (\%) & $H(gg)$ &  $H(WW)$ & $H(b\bar b)$ & $H(q\bar q)$ & $Z(q\bar q)$ & $ZZ(q\bar q)$\tabularnewline
\hline
\hline
 [0.02, 0.03] & $-5$  & $22$& $-3$ & $-0.3$ & $-0.4$ & $-0.4$\tabularnewline
\hline                      
 [0.05, 0.07] & $-2$  & $-9$& $3$  & $3$    & $4$ & $4$\tabularnewline
\hline
\end{tabular}

\caption{
 Estimated hadronization uncertainties of normalized distributions of thrust
 for two representative bins of $\tau$.
\label{tab:hadro}}
\end{table}

\section{CEPC}
A circular electron-positron collider has been proposed
recently with a center-of-mass energy of 250 GeV and a total integrated luminosity
of 5 ${\rm ab^{-1}}$~\cite{CEPC-SPPCStudyGroup:2015csa}. It can serve as a Higgs
factory with the dominant production channel being the associated production with a
$Z$ boson, with a total cross section of about 212 fb~\cite{Chen:2016zpw}. One great advantage of the
$e^+e^-$ collider is that the Higgs boson events can be selected by measuring
the recoil mass $m_{\rm recoil}$, e.g., for $ZH$ production with the $Z$ boson decay into a pair
of visible fermions $f\bar f$,
\begin{equation}
m_{\rm recoil}^2=s-2E_{f\bar f}\sqrt s+m_{f\bar f}^2,
\end{equation}
where $E_{f\bar f}$ and $m_{f\bar f}$ are the total energy and invariant mass of
the fermion pair. The recoil mass spectrum should present a sharp peak at the
Higgs boson mass. The Higgs boson events can be selected with a high signal
to background ratio independent of the decay modes of the Higgs boson. Using the kinematic
information of the recoil system, we can boost all decay products back to the rest
frame of the Higgs boson and measure the event shape distributions in that frame.

Table~\ref{tab:tot} summarizes the decay BRs of the hadronic decays of the SM Higgs
boson and the expected numbers of events at the CEPC through $ZH$ production, with
the $Z$ boson decaying into electron or muon pairs. As one can see, the $q\bar q$ (light quarks)
channel is negligible in the case of the SM Higgs boson. All the hadronic channels
in Table~\ref{tab:tot} contribute to the distribution of the event shapes.
We must carefully select the one that we are interested in, which is the $jj$
($gg$+$q\bar q$) channel. To suppress the heavy-quark contributions, one
can use flavor tagging of the heavy quarks,
$b$ and $c$, a technique which is well established at hadron and lepton
colliders~\cite{1512.01094}. It has been shown that, assuming an efficiency of 97.2\% for identification
of gluon or light quarks $j$, the misclassification rate of a $b$ or $c$ quark to $j$ at CEPC could reach
8.9\% and 40.7\% respectively~\cite{CEPC-SPPCStudyGroup:2015csa,CEPCtalk}. Since there
are two quarks/gluons from the decay, by requiring both of them untagged one can remove 99(84)\%
of the $b\bar b$($c \bar c$) background while only changing the signal $jj$ by 6\%.
There are also backgrounds from other SM processes, especially from the SM $Zq\bar q$ production,
which have a flat distribution in the recoil mass. After applying further selection cuts,
e.g., on recoil mass, dilepton invariant mass, and the polar angle of the Higgs 
boson, we estimate a total signal ($jj$) efficiency of
50\%~\cite{CEPC-SPPCStudyGroup:2015csa,Chen:2016zpw}. We assume a total $q\bar q$-like
background of 30\% of the signal rate from Higgs boson decays to $b\bar b$, $c\bar c$ and
the SM $Zq\bar q$ production of which about 10\% is from $b\bar b$ and $c\bar c$ as can
be calculated from the misidentification rates and various decay BRs. The normalization
of $Zq\bar q$ background is estimated according to Fig.~7 in Ref.~\cite{Chen:2016zpw}.
A second category of backgrounds are from decays to $WW^*$, $ZZ^*$ and further to four
quarks. Since they are away from the peak region of our signal, as shown in Fig.~\ref{fig:thrust},
they do not have
a large impact to the measurement of the light-quark couplings. We estimate
a total rate of 60\% of the signal for these four-quark backgrounds after all selection cuts.
They can be further suppressed if additional cuts on dijet invariant masses are used.
Noted we do not impose any selection cuts directly in our calculations of the signal and
backgrounds but rather estimate their effects on signal and background normalizations.     

\begin{table}[h!]
\centering
\begin{tabular}{c|cccccc}
\hline
$Z(l^+l^-)H(X)$ & $gg$ & $b\bar b$ & $c\bar c$ &  $WW^*(4h)$ & $ZZ^*(4h)$ & $q\bar q$\tabularnewline
\hline
\hline
$BR$ [\%] & $8.6$ & $57.7$ & $2.9$ & $9.5$ & $1.3$ & $\sim 0.02$\tabularnewline
\hline
$N_{event} $ & $6140$ & $41170$ & $2070$ & $6780$ & $930$ & $14$\tabularnewline
\hline
\end{tabular}

\caption{The decay branching ratios of the SM Higgs boson with a mass of
$125$ GeV to different hadronic channels~\cite{Heinemeyer:2013tqa} and the corresponding expected
numbers of events in $ZH$ production, with subsequent decays at a $e^+e^-$
collider with $\sqrt s={\rm 250\ GeV}$ and an integrated luminosity of 
5 ${\rm ab^{-1}}$. Only decays of the associated $Z$ boson
to electrons and muons are included. $h$ represents any of the quarks
except the top quark and $q$ are light quarks. 
\label{tab:tot}}
\end{table}

Including both the signal and backgrounds, the event shape distributions
at hadron level can be expressed as
\begin{align}\label{eq:eve}
\frac{d N}{d O}= & N_S(rf_{H(q\bar q)}(O)+(1-r)f_{H(gg)}(O))
     + N_{B,1}f_{H(b\bar b)}(O) \nonumber \\  
     &+ N_{B,2}f_{ZZ(q\bar q)}(O) + N_{B,3}f_{H(WW)}(O),
\end{align}
where $N_S$, $N_{B,1}$, $N_{B,2}$, and $N_{B,3}$ are the expected number of events
for the signal, the $q\bar q$-like backgrounds from heavy quarks in Higgs decay and from
$Zq\bar q$ production, and the four-quark background,
respectively.
The interference effects between the Higgs gluonic and fermionic
couplings from higher-orders in QCD are suppressed by an additional factor of quark mass over Higgs boson mass
due to chirality violation and are negligible here.
We normalize the signal rate to the SM result,
$N_S=\lambda N_{S,SM}$
with $\lambda=\sigma(HZ){\rm BR}(jj)/\sigma(HZ){\rm BR}(jj)_{SM}$.
From previous discussions, we have $N_{S,SM}=3070$ and
$N_{B,1(2,3)}=0.1(0.2,0.6)N_{S,SM}$. In addition, $r={\rm BR}(q\bar q)/{\rm BR(jj)}$
is the fraction of the Higgs boson BR to light quarks which we
would like to measure. Both $r$ and $\lambda$ allow possible
deviations from the SM which has $r=0$ and $\lambda=1$. 
Noted we assume the Higgs boson couplings to be SM-like when calculating
various backgrounds, except for the couplings to gluon and light quarks.
Thus the modification of the gluon coupling can only be due to top quark
or new colored particles in the loop. 
In Eq.~(\ref{eq:eve})
$f_{H(q\bar q)/(b\bar b)/(gg)/(WW)}$ are the normalized distributions of the Higgs boson decay
to light quarks, bottom quarks, gluons, or four quarks through $W$ boson pair as shown in
Fig.~\ref{fig:thrust}. 
$f_{ZZ(q\bar q)}$ is the normalized distribution for $Zq\bar q$ production.
We simply assume a shape of $f_{H(b\bar b)}$ for the heavy-quark components
of the backgrounds. 
Impact of using the actual mixture of bottom- and charm-quark distributions
are small. 

We take into account 11 independent systematic uncertainties for the thrust distribution.
Two of them are the perturbative uncertainties of the normalized distribution
$f_{H(gg)}$, as shown in Fig.~\ref{fig:thrust}. Each of them is (anti-)correlated
among all bins. We include five systematic errors for various normalized shapes
in Eq.~(\ref{eq:eve}) due to the hadronization uncertainties as discussed earlier.
The other four are for the normalization 
of the signal $N_S$ and of the backgrounds $N_{B,1}$, $N_{B,2}$, and $N_{B,3}$
in Eq.~(\ref{eq:eve}).
We do not assume any correlations among them. Normalization uncertainties on each of
the backgrounds are set to 4\%. Normalization of the signal can be measured separately
using hadronic decays of the $Z$ boson in $ZH$ production with the Higgs boson decay to $jj$,
and the uncertainty is estimated to be 3\%~\cite{CEPC-SPPCStudyGroup:2015csa}. 
We have not included any perturbative uncertainties for the normalized shapes of
$q\bar q$ signal and various backgrounds. We estimate their effects to be comparable
or smaller than those of hadronization uncertainties with future high precision
calculations.

\begin{figure}[!h]
  \begin{center}
  \includegraphics[width=0.45\textwidth]{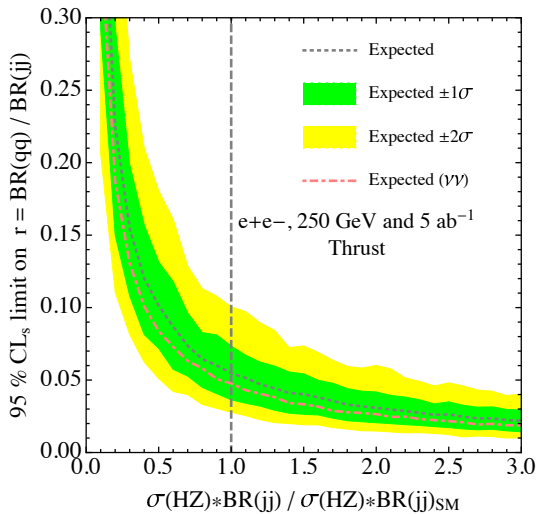}
  \end{center}
  \vspace{-2ex}
  \caption{\label{fig:ecl1}
  Expected 95\% $\rm CL_s$ exclusion limit on $r$ and the $1\sigma$ and $2\sigma$
  fluctuations as a function of the total cross section of the Higgs boson decay
  to $jj$ normalized to the SM value. The dot-dashed line is the expected exclusion limit
  when invisible decays of the $Z$ boson are also included in the analysis.}
\end{figure}

We study the expected exclusion limit on
$r$, as a function of $\lambda$, assuming the decay to $q\bar q$ vanishes. 
We generate a large ensemble of pseudo-data according to Eq.~(\ref{eq:eve}) with the hypothesis of $r=0$.
Systematic uncertainties are treated using nuisance parameters.
Statistical fluctuations are included according to Gaussian distributions based on the
expected event rates in each bin.
For each of the pseudo-data we determine the exclusion limit on $r$ by using
the profiled log-likelihood ratio $q_{\mu}$ as our
test-statistic~\cite{1007.1727} together with the ${\rm CL_s}$ method~\cite{Read:2000ru}.
Fig.~\ref{fig:ecl1} shows the expected 95\% ${\rm CL_s}$ exclusion limit on
$r$ (in the dashed line) from the thrust distribution. The colored bands indicate
the $1\sigma$ and $2\sigma$ fluctuations of the expected exclusion limit. In case
the true theory is the SM, the expected exclusion limit on $r$ can reach 0.056,
which is the intersection of the curve and the vertical line.
That corresponds to a decay BR of $0.48$\% to $q\bar q$. In term of the Yukawa
coupling strength, that implies $y_q < 0.091 y_b$ for any of $q=u,d,s$,
with $y_b$ being Yukawa coupling of the bottom quark in the SM.

\begin{table}[h!]
\centering
\begin{tabular}{l|cccc}
\hline
& no sys.&+pert.&+nor.&+had.\tabularnewline
\hline
limit on $r$ 
& $0.036$ & $0.040$ & $0.045$ & $0.056$  \tabularnewline
limit on $r$ (lumi.$\times 10^3$) 
& $0.0012$ & $0.0014$ & $0.018$ & $0.019$  \tabularnewline
\hline
\end{tabular}

\caption{
 Impact of various systematic uncertainties on the expected 95\% CL$_s$ exclusion limit
 of $r$ with $\lambda=1$ and a luminosity of 5 ${\rm ab}^{-1}$ or 5000 ${\rm ab}^{-1}$.
 Numbers correspond to the exclusion limit without any systematic errors, and adding
 various systematic errors in succession. 
\label{tab:sys}}
\end{table}

The sensitivity on $r$
can be understood as below. There
are two major discrimination powers when testing finite $r$ against the SM case.
One is from the $q\bar q$-peak region and the other is from the $gg$-peak region.
If neglecting statistical errors, in the $q\bar q$-peak
region, a finite $r$ (an enhancement) can only be mimic by a systematic shift
of $N_{B,1(2)}$. Thus the 95\% ${\rm CL_s}$ limit approximately corresponds
to $r\approx 0.3*0.04*1.64\approx 0.02$. On the other hand, in the $gg$-peak region,
a finite $r$ (a deficit) can only be compensated by a systematic shift
of $N_S$. The limit is about $r=0.03*1.64\approx0.05$. When combining
both the limit is better than 0.02. 
After considering the statistical fluctuations and other systematic errors the
limit increases to 0.056 as shown in Fig.~\ref{fig:ecl1}. We further illustrate
impact of various systematic uncertainties on the exclusion limit
of $r$ in Table~\ref{tab:sys}.
We show numbers correspond to the exclusion limit without any systematic errors,
and adding various systematic errors in succession.
We can see the uncertainty is dominated by the
statistical error. The hadronization uncertainties show a moderate
impact. For comparison we also list the results with a data sample of 1000 times
larger. 

We can also include invisible decays of the
associated $Z$ boson in the analysis. They have a total
rate 3 times larger than to electrons and muons and suffer from a relatively larger
$Zq\bar q$ background due to a degradation of the signal-background separation power from
the recoil mass. Thus, we simply assume that once the $\nu\nu$ channels are included, both the
signal and backgrounds will double. The expected limit is again plotted in Fig.~\ref{fig:ecl1},
which can reach 0.047 with the SM assumption.

In principle, several of the backgrounds, e.g. $f_{H(b\bar b)}$ and $f_{ZZ(q\bar q)}$
can be measured directly in a controlled region from independent data sample.
We briefly comment on the possibilities in below.
\begin{itemize}
\item
Heavy-quark components: in this case one can require the quark/gluon being flavor
tagged rather than untagged. With a typical $b$-tagging efficiency of 60\%, the
misidentification rate for light flavors are negligible~\cite{1512.01094} not mentioning further
suppression from the Higgs boson decay branching ratio. Thus we can arrive at 
a pure sample of $b\bar b$ of around $2\times 10^4$ events for CEPC. That corresponds to an
uncertainty of 1.5\% from statistical fluctuations for the bin $[0.02,\,0.03]$ of $\tau$,
comparable to the number in
Table~\ref{tab:hadro}. One question needs to
be addressed is how various flavor-tagging algorithm may change distributions of the
event shape observables.  
\item
$Zq\bar q$ component: we can require the recoil mass of the lepton pair to be slightly
off the Higgs boson mass to remove all events from Higgs boson decay. For instance we
can select two recoil mass windows of $[110,\,120]$ GeV and $[130,\,140]$ GeV and take
the average of the two distributions measured as $f_{ZZ(q\bar q)}$. That contains about
$4\times 10^3$ events in each window and gives an uncertainty of 2.1\% for the 
bin $[0.02,\,0.03]$ of $\tau$, which is much larger than the number in Table~\ref{tab:hadro}.  
\end{itemize}

\begin{figure}[!h]
  \begin{center}
  \includegraphics[width=0.45\textwidth]{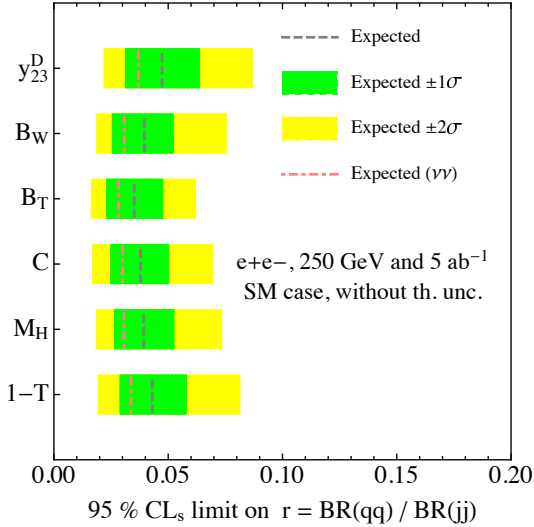}
  \end{center}
  \vspace{-2ex}
  \caption{\label{fig:ecl2}
  Expected 95\% $\rm CL_s$ exclusion limit on $r$ and the $1\sigma$ and $2\sigma$
  fluctuations based on measurements of different event shape observables and assuming
  a theory of the SM. Theoretical uncertainties on the event shape distributions are not included.}
\end{figure}

Similar exclusion limits can be set based on other event shape observables which are
summarized in Fig.~\ref{fig:ecl2} for $\lambda=1$. Definitions of the event shape observables
shown in Fig.~\ref{fig:ecl2} can be found in Refs.~\cite{Abbiendi:2004qz,Heister:2003aj}.
Here, only the statistical error and the
systematic uncertainty on the signal and background normalizations are included in the analysis.
Thus the limits shown here are optimistic concerning various theoretical uncertainties.  
As already seen in Table~\ref{tab:sys} various theoretical uncertainties contribute
equally as the statistical uncertainty for the thrust distribution. 
The binnings used in the analysis for all other
distributions are chosen to be the same as in Ref.~\cite{hep-ph/0503051}. All distributions
show a similar sensitivity to the light-quark Yukawa couplings.\\

\section{Discussion and summary}
It is interesting to compare our sensitivity to the light-quark
Yukawa couplings with the projection of
the LHC and HL-LHC. Ref.~\cite{1606.09621} claims an expected 95\% CL limit of
the Yukawa couplings $y_{u,d}<0.4y_b$, for LHC 13 TeV run with a total luminosity
of 300 ${\rm fb^{-1}}$, based on analyzing the $p_T$ distribution of the Higgs
boson. Ref.~\cite{1606.09253} reports a sensitivity of $y_s\sim 0.52 y_b$
for the strange quark at the HL-LHC.
Comparing with results above, our method provides a much stronger
sensitivity of $y_{u,d,s}<0.091y_b$ (95\% $\rm CL_s$).
The major limitation on
probing the light-quark Yukawa couplings at the LHC/HL-LHC is that the $gg$ parton
luminosity is much larger than the $q\bar q$ ones for a Higgs boson mass
of 125 GeV. Thus, a small downward shift of the $gg$ induced cross sections comparing
to experimental data, either due to the experimental or theoretical uncertainties, can allow for a
much larger light-quark Yukawa coupling. 

We also comment on the comparison of our proposal with the possibility of using
gluon/quark jet discriminators. On the theory side, the event shape distributions can
be calculated systematically in perturbative QCD, and the theoretical uncertainties
are under control. Experimentally, the hadronic even-shape observables have
been studied extensively at LEP. The experimental systematics are well understood.
By comparing with the experimental results on the $\alpha_s(M_Z)$ measurement~\cite{hep-ph/0503051,1101.1470},
we found the sensitivity obtained in this study is realistic. Even
after all the experimental systematics are included, the expected exclusion limit
should not change greatly. 

In summary, we have proposed a novel idea for measuring the light-quark Yukawa 
couplings using hadronic event shape distributions in addition
to the conventional measurement of Higgs couplings at lepton colliders. We
show that for a $e^+e^-$ collider with a center-of-mass energy of 250 $\rm GeV$
and an integrated luminosity of 5 $\rm ab^{-1}$ one can expect to exclude a decay BR
of 0.48\% for the Higgs boson decay to $q\bar q$, at 95\% $\rm CL_s$, with $q$
be any of the $u,d,s$ quarks, assuming a hypothesis of SM-like theory and only modifications
to the Higgs boson couplings to gluon and light quarks. That corresponds
to an exclusion limit on a light-quark Yukawa coupling of about 9\% of the strength
of the bottom quark coupling in the SM.

\begin{acknowledgments}
JG would like to thank Manqi Ruan, Hua Xing Zhu, C. Wagner and E. Berger
for useful conversations, and
J. Huston for proofreading of the paper.
The work of JG is sponsored by Shanghai Pujiang Program.
\end{acknowledgments}

\end{document}